\documentclass[epj]{svjour}
\usepackage{graphics}
\begin{document}

\title{Refraction of light by light in vacuum}

\author{X. Sarazin\inst{1}\thanks{sarazin@lal.in2p3.fr}, F. Couchot\inst{1}, A. Djannati-Atai\inst{2}, O. Guilbaud\inst{3}, S. Kazamias\inst{3}, M. Pittman\inst{4}, M. Urban\inst{1}
}

\institute{LAL, IN2P3/CNRS, Universit\'e Paris-Saclay, Orsay, France
 \and APC, IN2P3/CNRS, Univ Paris Diderot, Paris, France
 \and LPGP, Universit\'e Paris-Saclay, Orsay, France
 \and LUMAT, Universit\'e Paris-Saclay, Orsay, France}

\titlerunning{Refraction of light by light in vacuum}
\authorrunning{X. Sarazin {\it et al.}}
\date{Received: / Revised version: }
%
\abstract{
In very intense electromagnetic fields, the vacuum refractive index is expected to be modified due to nonlinear quantum electrodynamics (QED) properties.
Several experimental tests using high intensity lasers have been proposed to observe electromagnetic nonlinearities in vacuum, such as the diffraction or the reflection of intense laser pulses. 
We propose a new approach which consists in observing the refraction, i.e. the rotation of the waveplanes of a probe laser pulse crossing a transverse vacuum index gradient. The latter is produced by the interaction of two very intense and ultra short laser pulses, used as pump pulses. 
At the maximum of the index gradient, the refraction angle of the probe pulse is estimated to be $0.2 \times (\frac{w_0}{10 \mathrm{\mu m}})^{-3} \times \frac{I}{1 \mathrm{J}}$~picoradians, where $I$ is the total energy of the two pump pulses and $w_0$ is the minimum waist (fwhm) at the interaction area.
Assuming the most intense laser pulses attainable by the LASERIX facility ($I = 25$~J, 30~fs fwhm duration, 800~nm central wavelength) and assuming a minimum waist of $w=10 \mathrm{\mu m}$ (fwhm) (corresponding to an intensity of the order of $10^{21}$~W/cm$^2$), the expected maximum refraction angle is about 5~picoradians. 
An experimental setup, using a Sagnac interferometer, is proposed to perform this measurement.
} 

\maketitle

\section{Introduction}

When the vacuum is stressed by very intense electromagnetic fields, it should behave as a nonlinear optical medium, as predicted initially by Euler-Heisenberg~\cite{euler-heisenberg}, and reformulated later as nonlinear QED interaction.  
So far, electromagnetic nonlinearity in vacuum has never been observed at low energy, i.e. on macroscopic scale with light beams. 
Several methods have been developed to detect it: vacuum birefringence induced by an external magnetic field, frequency doubling (parametric amplification) or photon fusion induced by an external field, and direct photon-photon scattering (see e.g., the review in~\cite{battesti-rizzo}). 
The most sensitive experimental test today is the search for vacuum magnetic linear birefringence in the presence of an external magnetic field~\cite{bmv}\cite{pvlas}. 
No signal has yet been observed and the best achieved limit at 90\% confidence level, obtained with permanent rotating magnets, is about two orders of magnitude above the QED prediction~\cite{pvlas}.

In the last years, there has been an increasing interest in the search for electromagnetic nonlinearity in vacuum by measuring the scattering of intense laser pulses. 
At the interaction area of two laser pulses, the nonlinear quantum coupling of the intense electromagnetic fields produces an increase of the vacuum refractive index. The optical index imprinted on the vacuum induces the scattering of a probe beam (the pump beam itself or an external probe beam). 
Several  configurations have been studied. We can mention as examples the diffraction of a high-intensity laser pulse which crosses another counter propagating high-intensity laser pulse with smaller diameter~\cite{tommasini}~\cite{king}, the diffraction of a probe beam by a standing electromagnetic wave formed by two counter propagating intense laser pulses~\cite{dipiazza}, and recently the reflection of a probe beam on the interaction area of pump laser pulses~\cite{gies}. 

In all these proposals, the scattering of the laser pulse is similar to the classical diffraction or reflection of a light beam by an optical aperture. 
Here the optical aperture corresponds to the vacuum area where the optical index is enhanced by the interaction of two intense laser pulses. It is worthwhile noting that the diffraction and reflection occur, even if the transverse profile of the refractive index is constant. 
However, when a light beam crosses an optical aperture with a transversely varying index, the light beam is also refracted: the beam is fully bended towards regions of higher index because of the rotation of the wave planes. 

Here we propose to study this refraction effect. 
The vacuum refractive index induced by the interaction of two laser pulses is proportional to the intensity of the pulses. 
Since the transverse intensity profile is non uniform, it induces a transversal gradient of the vacuum index. 
Therefore the phase velocities in the laser pulse are gradually slower close to the interaction area, and the wave planes are gradually rotated. 
The laser pulse is hence refracted. 
In the scattering processes, only a very small fraction of photons are diffracted or reflected, although the deflection angle is relatively high. 
Here, in the refraction effect, the whole beam undergoes bending but the refraction angle is tiny.

\begin{figure}\sidecaption
\resizebox{1.\hsize}{!}  {\includegraphics*{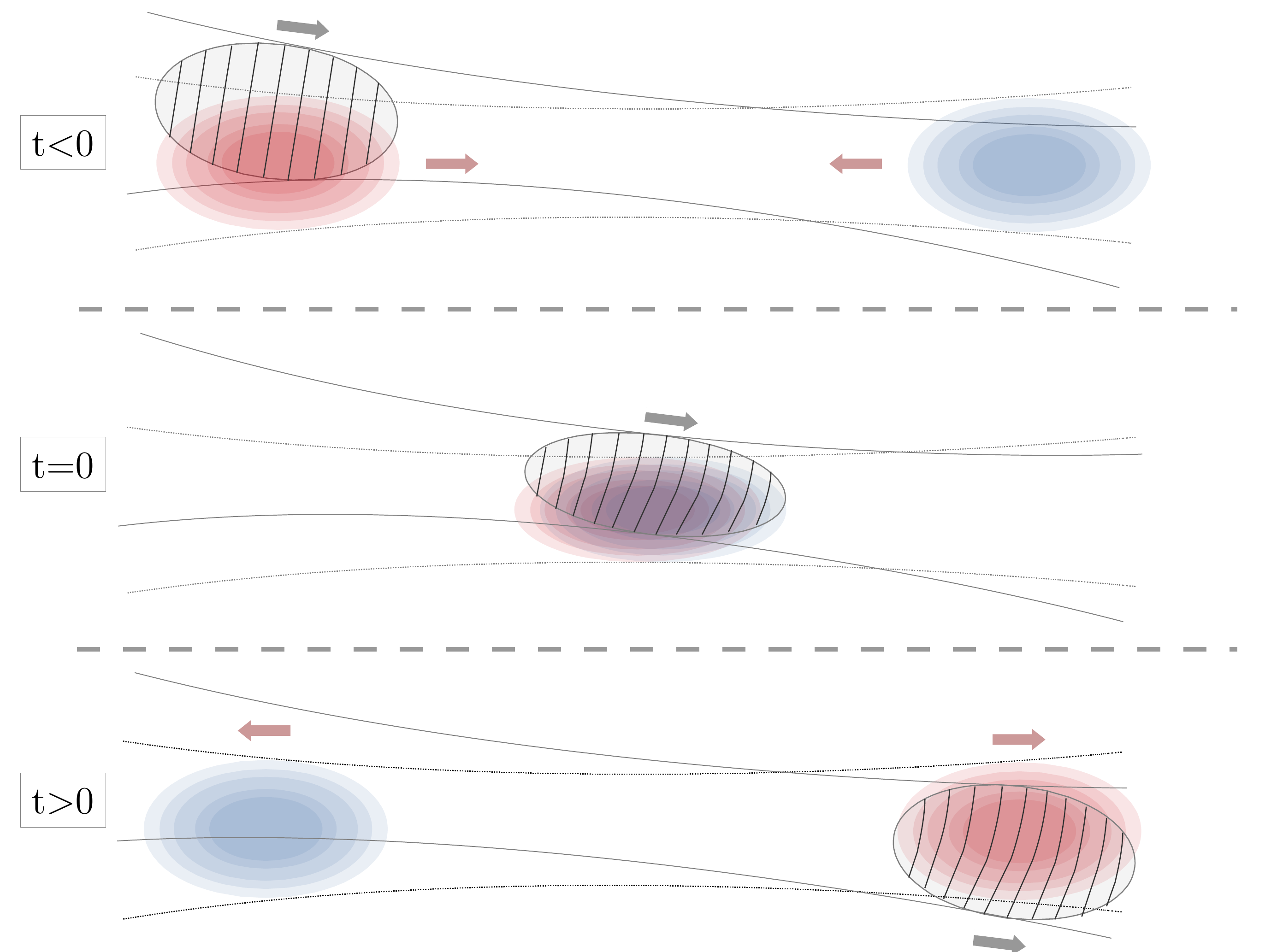}}
  \caption{Schematic view of the interaction of the counter-propagating pump pulses (red and blue disks) and the refraction of the probe pulse (disk with line). The lines inside the probe pulse correspond to the wave planes gradually rotated by the vacuum index gradient. They have been illustrated in a simplified way. Here the propagation axis of the probe pulse is slightly tilted and transversely shifted.}
  \label{fig:schema}
\end{figure}

The refraction of a light beam in a vacuum stressed by intense electromagnetic fields has been experimentaly tested only once in 1960 by R.V. Jones~\cite{jones} who looked for the deviation of a light beam passing through a transverse static magnetic field shaped to mimic an optical prism. The magnetic field strength was about 1~T and results indicated that the deflection of the light beam was less than 0.5~picoradians. Although the sensitivity of the experiment was incredibly high given the available technology at the time, this limit was almost 10 orders of magnitude above the expected signal from Euler-Heisenberg model, because of a relatively low magnetic field strength. 
We propose to develop an experiment, similar to that of R.V. Jones, with a much stronger transient magnetic field produced by the interaction of two counter propagating intense laser pulses. 
Since the measurement of the wave planes rotation within very intense laser pulses seems experimentaly unrealistic, we propose to measure the refraction by using a third laser pulse (a probe pulse) of lower intensity, as illustrated in Figure~\ref{fig:schema}.

This paper is organized as follows: we first outline the nonlinear properties of the electromagnetic quantum vacuum. 
We then show how the vacuum refractive index is modified at the intersection of two counter propagating laser pulses, and calculate the transverse index gradient. 
The rotation angle of the wave planes (the refraction) is then computed numerically for a probe beam crossing this index gradient.
Finaly, a possible measurement of this refraction with a Sagnac interferometer is discussed.

In the following, we use the caracteristics of the laser pulses delivered by the LASERIX facility~\cite{laserix}: duration of the pulses of $\Delta t = 30$~fs (fwhm), 800~nm central wavelength, a total energy of the two pump pulses of 25~J (12.5~J for each) and a repetition rate of 0.1~Hz. A minimum waist of the pump pulses at the interaction of $\Delta_{\bot} = 10 \mathrm{\mu m}$ (fwhm) is also assumed.

\section{Nonlinear properties of the electromagnetic quantum vacuum}
\label{sec:nlqed}

In classical electrodynamics, the  electromagnetic properties of the vacuum are described by two fundamental constants: the vacuum permittivity $\epsilon_0$ and the vacuum permeability $\mu_0$. 
In the presence of an eletric field $\mathbf{E}$, the induced vacuum polarization $\mathbf{D}$ is given by 
\begin{eqnarray}
\mathbf{D}=\epsilon_0 \mathbf{E}
\end{eqnarray}
and in the presence of a magnetic field $\mathbf{H}$, the magnetic induction field $\mathbf{B}$ is given by
\begin{eqnarray}
\mathbf{B}= {\mu_0} \mathbf{H}
\end{eqnarray}
These two constants define the phase velocity of an electromagnetic wave propagating in vacuum by the relation $c=1/\sqrt{\epsilon_0 \mu_0}$. 

In a standard material medium, the light velocity is decreased because of a modification of the permittivity and permeability, given by
\begin{eqnarray}
\label{eq:chi-medium}
\nonumber 
\mathbf{D} & = & \epsilon \mathbf{E} = \epsilon_0 (1+\chi_e) \mathbf{E} = \epsilon_0 \mathbf{E} +  \mathbf{P} \\
\mathbf{B} & = & \mu \mathbf{H} = \mu_0 (1+\chi_m) \mathbf{H} = \mu_0 \mathbf{H} + \mu_0 \mathbf{M}
\end{eqnarray}
where $\chi_e$ and $\chi_m$ are the electric and magnetic susceptibilities of the medium, $\mathbf{P}$ is the electric polarization
\begin{eqnarray}
\label{eq:polarization}
\mathbf{P}=\epsilon_0 \chi_e \mathbf{E}
\end{eqnarray}
and $\mathbf{M}$ is the  magnetization
\begin{eqnarray}
\label{eq:magnetization}
\mathbf{M} = \chi_m \mathbf{H} = \frac{1}{\mu_0} \frac{\chi_m}{1+\chi_m} \mathbf{B}
\end{eqnarray}
The optical refractive index is then given by 
$$n=\frac{\sqrt{\epsilon \mu}}{\sqrt{\epsilon_0 \mu_0}} = \sqrt{(1+\chi_e)(1+\chi_m)}$$
It is well established that in a high field regime, material media become optically nonlinear: the susceptibilities are not constant but depend on the magnetic and electric fields applied to this medium.

Since the early work by Heisenberg and Euler~\cite{euler-heisenberg}, it is predicted that the electromagnetic vacuum must also behave as a nonlinear optical medium in presence of very intense electromagnetic fields. 
The virtual electron-positron fluctuations which are produced in the QED vacuum, are expected to induce nonlinear couplings between real electromagnetic fields. This modifies the interactions and propagations of the fields and leads to modified Maxwell equations with a vacuum polarisation (similar to the ones in a medium) given, at the lowest orders in the fields, by (see e.g., Refs~\cite{battesti-rizzo} and \cite{soljacic})
\begin{eqnarray}
\label{eq:polarisations-HE-1}
\nonumber\mathbf{P} & = & \xi \epsilon_0^2 \left[2 (E^2-c^2B^2) \mathbf{E}+7 (\mathbf{E}\cdot\mathbf{B}) c^2\mathbf{B} \right]\\
\mathbf{M}          & = & -\xi \epsilon_0^2 c^2  \left[2 (E^2-c^2B^2) \mathbf{B}-7 (\mathbf{E}\cdot\mathbf{B}) \mathbf{E} \right]
\end{eqnarray}
with
\begin{eqnarray}
\label{eq:xi}
\xi=\frac{4 \alpha^2\hbar^3}{45 m_e^4 c^5} \simeq 3.35 \ 10^{-30} \ \mathrm{\frac{m^{3}}{J}}
\end{eqnarray}

Because of this polarisation, the optical refractive index of the vacuum, for a light propagating in the presence of an external electromagnetic field, is no longer equal to 1 but depends on the external field amplitude.

There are also QED corrections that include the spatial and temporal derivatives of the fields $\mathbf{E}$ and $\mathbf{B}$~\cite{mamaev}. However, as discussed in~\cite{soljacic}, the derivative corrections are important when the wavelength of the light $\lambda$ is comparable to the Compton wavelength $\lambda_C$ of the electron, while the Heisenberg-Euler correction given in Equation~\ref{eq:polarisations-HE-1} is important when the electric field becomes comparable to the critical electric field $E_c=m^2c^3/(e\pi\hbar)$. 
More generaly, it is shown in~\cite{soljacic} that the derivative corrections are negligible as long as $(\lambda_C/\lambda)^2 \ll (4/\pi^3)(E/E_c)^2$. Since we are studying a regime where the derivative corrections are negligible, we adopt the locally constant field approximation and we consider only the Heisenberg-Euler correction given by Equation~\ref{eq:polarisations-HE-1}.

\section{Refraction of a probe pulse by the interaction of two pump pulses}

We study the configuration of two laser pulses ($\mathbf{E}_1$,$\mathbf{B}_1$) and ($\mathbf{E}_2$,$\mathbf{B}_2$) with the same intensity and orthogonal linear polarizations. These two pulses, named pump pulses, propagate towards opposite directions along the same axis. They cross each other at focus. The coupling of the two fields induces an increase of the vacuum refractive index which is proportional to the intensity of the fields. 
Since the transverse intensity profile of the pump pulses is not uniform, a gradient of the vacuum index occurs in the transverse direction of the propagation. 
A third pulse of lower energy (in the millijoule range), used as a probe pulse, propagates simultaneously to one of the pump pulses, as illustrated in Figure~\ref{fig:schema}, and crosses the transverse index gradient imprinted on the vacuum. The phase velocities in the probe pulse are gradually slower close to the interaction area, and the wave planes are gradually rotated. 

In the following, we first calculate the gradient of the vacuum index induced by the interaction of the two pump pulses. We then calculate numerically the rotation angles (the refraction angles) of the wave planes of the probe pulse, after having crossed the interaction area.

\subsection{Variation of the vacuum refractive index induced by two counter propagating laser pulses}

In the interaction area, the electromagnetic fields of each pump pulse (labeled 1 and 2) are chosen to be two gaussian wave packets of same amplitude:
\begin{eqnarray}
\label{eq:paquet1}
\nonumber \mathbf{E_1} & = & A(x,y) F(z,t) \cos[\omega (t-z/c) -\phi_1] \ \mathbf{u}_x\\
          \mathbf{B_1} & = & \frac{A(x,y)}{c} F(z,t) \cos[\omega (t-z/c) -\phi_1] \ \mathbf{u}_y
\end{eqnarray}
\begin{eqnarray}
\label{eq:paquet2}
\nonumber \mathbf{E_2} & = & A(x,y) F(z,-t) \cos[\omega (t+z/c) -\phi_2] \ \mathbf{u}_y\\
          \mathbf{B_2} & = & \frac{A(x,y)}{c} F(z,-t) \cos[\omega (t+z/c) -\phi_2] \ \mathbf{u}_x
\end{eqnarray}
with
\begin{eqnarray}
\label{eq:simple}
A(x,y) & = & A_0 \ e^{-(2\ln\!2 \times (x^2+y^2)/\Delta_{\bot}^2)} \\
F(z,t) & = & e^{-(2 \ln\!2 \times (t-z/c)^2/\Delta t^2)}
\end{eqnarray}
where $z$ is the axis of propagation, $x$ and $y$ are the transverse axes, $\mathbf{u}_x$ and $\mathbf{u}_y$ are the unit vectors along $x$ and $y$ respectively, $\Delta_{\bot}$ is the width (fwhm) of the transverse intensity profile, $\Delta t$ is the time width (fwhm) of the longitudinal intensity profile, and $t=0$ is arbitrarily defined when the two pulses entirely overlap. 
For simplicity, we will assume the phases $\phi_1$ and $\phi_2$ equal to 0 in the following. 
Experimentaly, the two pump pulses have to be sligthly tilted to prevent dangerous reinjection in the laser amplifiers. However, we assume here that the tilt angle between the two beam axes is small enough to be negligible as a first approximation. 
The transverse size, $\Delta_{\bot}$, of the pulses is also assumed to be constant at the interaction point and equal to the minimum waist. 
Indeed, the diffraction effects can be neglected, as a first approximation, for laser pulses of duration lower than few hundreds of femtoseconds and with a minimum waist larger than about 5~$\mathrm{\mu}$m. 

The amplitude $A_0$ of the electromagnetic wave packets depends on the total intensity $I$ of the two pump pulses, which is given by
\begin{eqnarray}
\label{eq:energie}
  I   =  2 \int ( \frac{\epsilon_0 E_1^2}{2} + \frac{B_1^2}{2\mu_0} ) dxdydz = 2 \epsilon_0 \int E_1^2 dxdydz 
\end{eqnarray}
At $t=0$, we get
\begin{eqnarray}
\label{eq:energie-t0}
\nonumber I  & = & 2 \epsilon_0 A_0^2 \int\!\!\!\int_{-\infty}^{+\infty}\!\! e^{-(4\ln2 (x^2+y^2)/\Delta_{\bot}^2)} dxdy\\
\nonumber    &   & \times \int_{-\infty}^{+\infty}\!\! e^{-(4 \ln2 z^2/(c\Delta t)^2)} \left(\cos(2\omega z/c)+1\right) dz \\
             & = & 2 \epsilon_0 A_0^2 \ \frac{\pi\Delta_{\bot}^2}{4\ln2} \ \frac{\sqrt{\pi} c \Delta t}{2\sqrt{\ln2}} \ \left( 1 + e^{-\omega^2 \Delta t^2 / (4\ln2)}  \right)
\end{eqnarray}
With femtosecond laser pulses at 800~nm central wavelength, $e^{-\omega^2 \Delta t^2 / (4\ln2)} \ll 1$ and
\begin{eqnarray}
\label{eq:energie-laserix}
A_0^2 = \frac{4\ln2 \ \sqrt{\ln2}}{\epsilon_0 \pi \sqrt{\pi} c} \frac{I}{ \Delta_{\bot}^2 \Delta t}   
\end{eqnarray}
Assuming a waist of the laser beam at the interaction area on the order of 10~$\mu$m (fwhm), and assuming a pulse duration of $\Delta t=30$~fs, the amplitude $A_0$ of the electric field of each pulse is $A_0 = 7.2 \ 10^{12} \sqrt{I/(1 \ \mathrm{J})}$~V/m. This corresponds to an amplitude of the magnetic field of $2.4 \ 10^{4} \sqrt{I/(1 \ \mathrm{J})}$~Tesla.

The sum of the two pulses is given by
\begin{eqnarray}
\nonumber \mathbf{E} & = & A(x,y) [ F(z,t) \cos(\omega (t-z/c)) \ \mathbf{u}_x \\
\nonumber            &   & + F(z,-t) \cos(\omega (t+z/c)) \ \mathbf{u}_y ] \\
\nonumber \mathbf{B} & = & \frac{A(x,y)}{c} [ F(z,-t) \cos(\omega (t+z/c)) \ \mathbf{u}_x \\
                     &   & + F(z,t) \cos(\omega (t-z/c)) \ \mathbf{u}_y ]
\end{eqnarray}
The fields are symmetric by inversion of $x$ and $y$. Consequently we have $\mathbf{E}^2=c^2 \mathbf{B}^2$ and the nonlinear vacuum polarizations given by Eq.~\ref{eq:polarisations-HE-1} become
\begin{eqnarray}
\label{eq:polarisations-HE-2}
\nonumber\mathbf{P} & = & 7 \xi \epsilon_0^2 c^2 \ (\mathbf{E}\cdot\mathbf{B}) \ \mathbf{B}\\
\mathbf{M}          & = & 7 \xi \epsilon_0^2 c^2  \ (\mathbf{E}\cdot\mathbf{B}) \ \mathbf{E}
\end{eqnarray}
By analogy with the susceptibilities of a medium given in Eq.~\ref{eq:polarization} and \ref{eq:magnetization}, we can define the electric and magnetic susceptibilities $\chi_{e0}$ and $\chi_{m0}$ of the electromagnetic vacuum, as
\begin{eqnarray}
\label{eq:chi-vacuum-def}
\nonumber\mathbf{P}\cdot\mathbf{E} & = &\epsilon_0\chi_{e0} E^2 \\
\mathbf{M} \cdot\mathbf{B} & = & \frac{1}{\mu_0} \frac{\chi_{m0}}{(1+\chi_{m0})} B^2 
\end{eqnarray}
Combining Eq.~\ref{eq:polarisations-HE-2} and \ref{eq:chi-vacuum-def}, and at the first order ($\chi_{e0} \ll 1$ and $\chi_{m0} \ll 1$), one gets
\begin{eqnarray}
\label{eq:chi-vacuum}
\nonumber
\chi_{e0} & = & 7 \xi \epsilon_0 c^2 \frac{(\mathbf{E}\cdot\mathbf{B})^2}{E^2} \\
\chi_{m0} & = & 7 \xi \epsilon_0 \frac{ (\mathbf{E}\cdot\mathbf{B})^2}{B^2} = \chi_{e0}
\end{eqnarray}
The emergence of these susceptibilities leads to a variation of the vacuum refractive index $1+\delta n_{\mathrm{QED}}$ inside the laser pulses, given at the first order in $\chi_{e0}$ and $\chi_{m0}$, by
\begin{eqnarray}
\label{eq:deltan}
\delta n_{\mathrm{QED}}(x,y,z,t) = 7 \xi \epsilon_0 c^2 \frac{(\mathbf{E}(x,y,z,t)\cdot\mathbf{B}(x,y,z,t))^2}{E^2(x,y,z,t)}
\end{eqnarray}

We can verify {\it a posteriori} that the increase of the vacuum refractive index is maximal when the two counter-propagating pump fields ($\mathbf{E}_1$,$\mathbf{B}_1$) and ($\mathbf{E}_2$,$\mathbf{B}_2$) have the same intensity. This is why we propose to use that configuration. As mentioned in the introduction, since the measurement of the waveplanes rotation within the intense pump pulses is experimentaly unrealistic, we use a third probe pulse with a much lower intensity to probe the vacuum index gradient.

\subsection{Calculation of the refraction angles of a probe beam}

We assume that the probe pulse propagates at the same time, parallel to one of the pump pulses. 
The intensity of the probe pulse (in the millijoule range) is about four orders of magnitude smaller than that of the pump pulses. Therefore its contribution to the index variation is neglected. 
The refraction of the probe pulse is a pure wave refraction effect. We can model it as the propagation of light rays (ray tracing) through the gradient of refractive index in the transverse direction of propagation. 

All calculations are done assuming the performance of the LASERIX facility, as given previously.

The calculation is performed numerically as follows.
We first sample the pump pulse $(\mathbf{E_1},\mathbf{B_1})$ in $x$, $y$ and $z$.
We then calculate at each point $x_i, y_i, z_i$ of the pulse, the total index gradient $\partial \Delta n(x_i,y_i,z_i)/\partial r$ in the transverse direction $\mathbf{r}=x_i \mathbf{u}_x + y_i \mathbf{u}_x$, integrated along the full crossing of the pulse with the second pump pulse $(\mathbf{E_2},\mathbf{B_2})$. 
Technically, since the terms of $\delta n_{\mathrm{QED}}$ in the transverse directions $(x,y)$ and in the propagation direction $(z,t)$ are independant, we first calculate the integrated index $\Delta n(x_i,y_i,z_i)$ by integrating $\delta n_{\mathrm{QED}}$ from $t=-\Delta t$ to $t=+\Delta t$. 
Then we calculate its gradient in the transverse direction $\mathbf{r}$, given by
\begin{eqnarray}
\label{eq:gradient_int}
\frac{\partial \Delta n(x_i,y_i,z_i)}{\partial r} = \frac{\partial}{\partial r} \left(\int_{-\Delta t}^{+\Delta t}{\delta n_{\mathrm{QED}}(x_i,y_i,z_i,t) dt}\right)
\end{eqnarray}
For very small deflection, the integrated refraction angle $\theta_r(x_i,y_i,z_i)$ in the transverse direction $\mathbf{r}$ is then given for each point $(x_i,y_i,z_i)$ of the pulse by
\begin{eqnarray}
\label{eq:angle}
\theta_r(x_i,y_i,z_i) = 2 c \Delta t  \ \frac{\partial \Delta n(x_i,y_i,z_i)}{\partial r} 
\end{eqnarray}

\begin{figure}\sidecaption
\resizebox{1.\hsize}{!}  {\includegraphics*{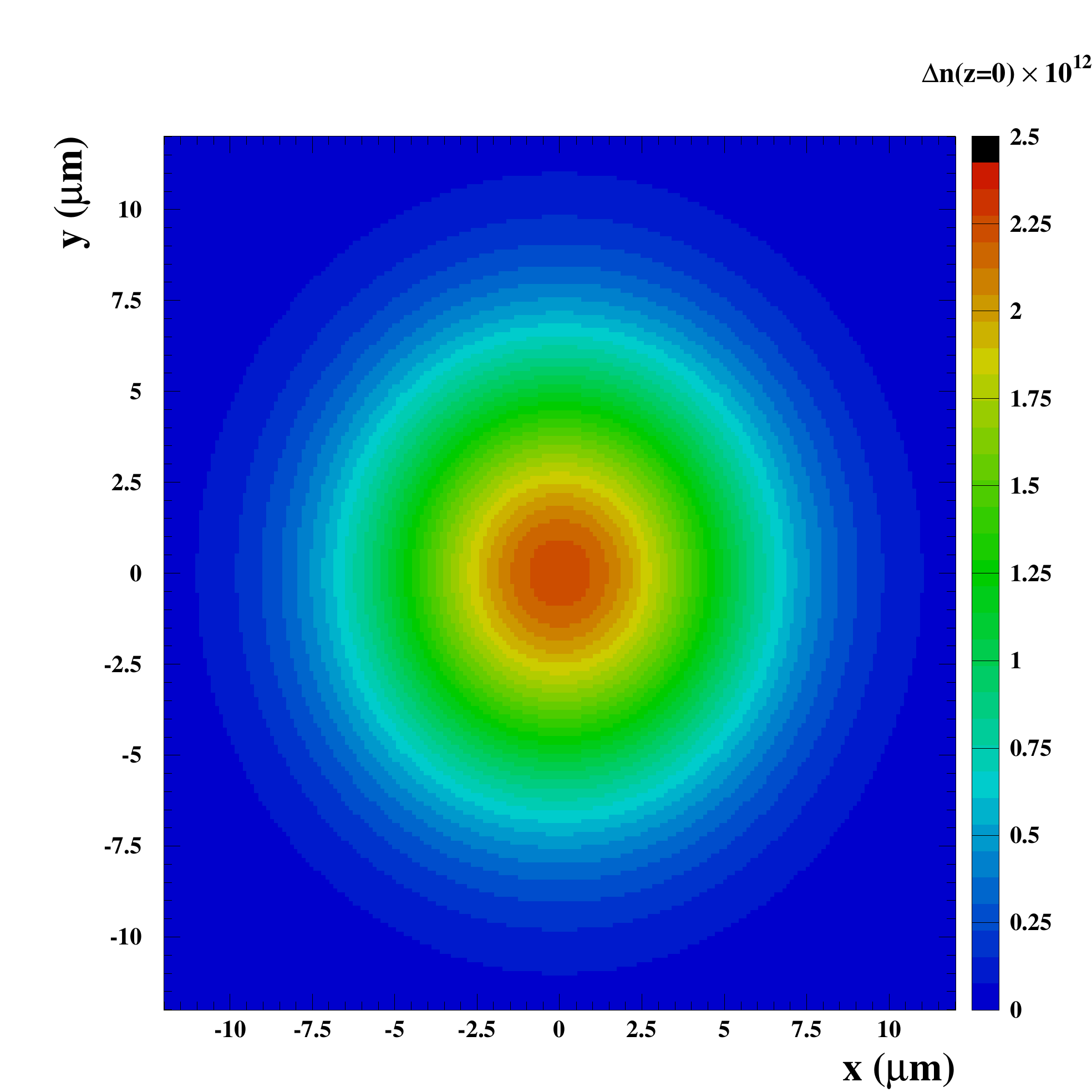}}
  \caption{The integrated index $\Delta n(x,y,z=0)$ ($\times 10^{12}$) calculated in the transverse plane $z=0$, as a function of the transverse coordinates $x$ and $y$. The total energy of the two pump pulses is 25~J, the duration of each pulse is 30~fs (fwhm) and the minimum waist is 10~$\mu$m at the focus, corresponding to an intensity of the order of $10^{21}$~W/cm$^2$.}
  \label{fig:index-xy}
\end{figure}

\begin{figure}\sidecaption
\resizebox{1.\hsize}{!}  {\includegraphics*{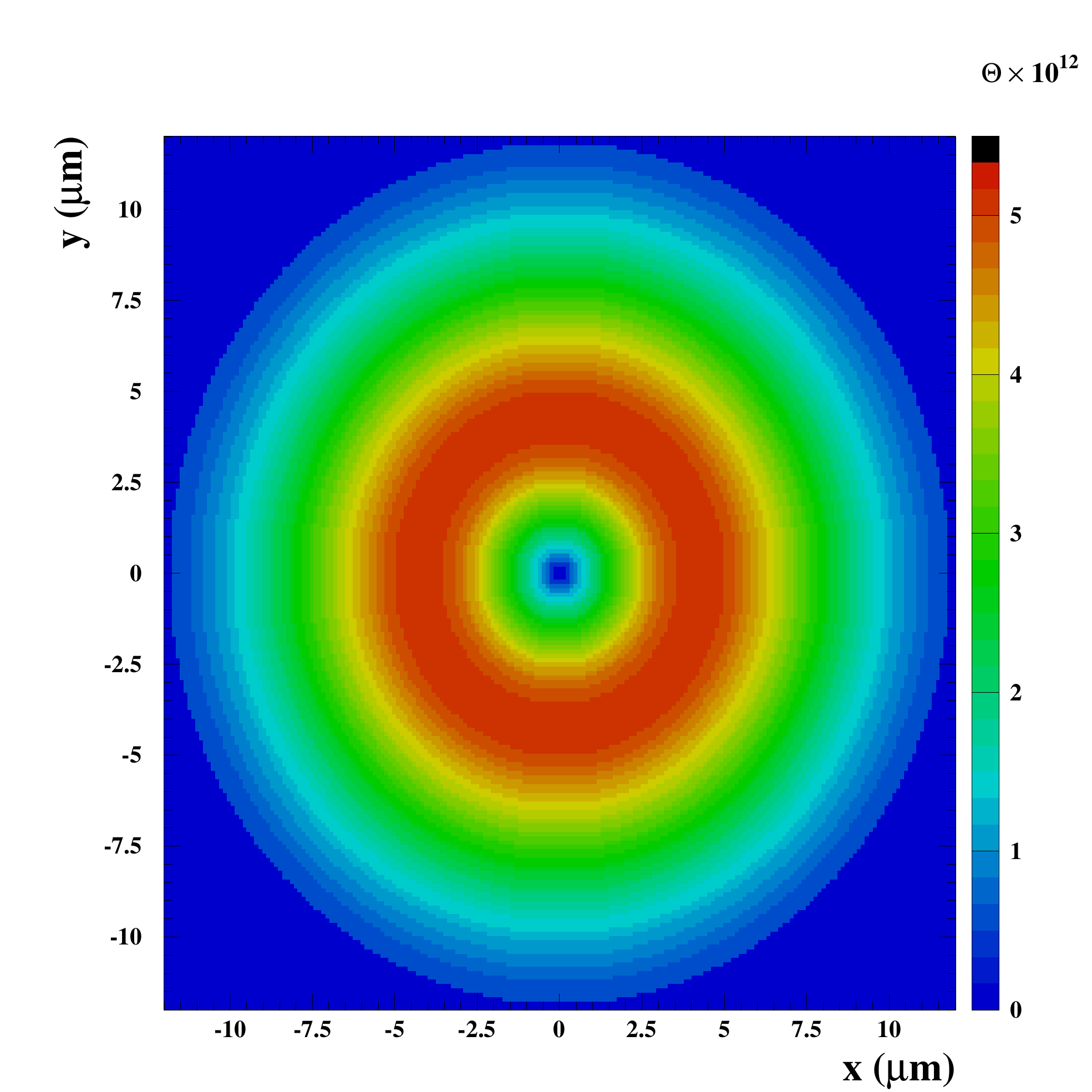}}
  \caption{The refraction angle $\theta_r(x,y,z=0)$ ($\times 10^{12}$ radians) calculated in the transverse plane $z=0$, as a function of the transverse coordinates $x$ and $y$. At the maximum of the index gradient ($r \simeq 5\mu$m), the refraction angle is about $5 \ 10^{-12}$~rad. Computation is done with same conditions as for Fig.~\ref{fig:index-xy}.}
  \label{fig:angle}
\end{figure}

\begin{figure}\sidecaption
\resizebox{1.\hsize}{!}  {\includegraphics*{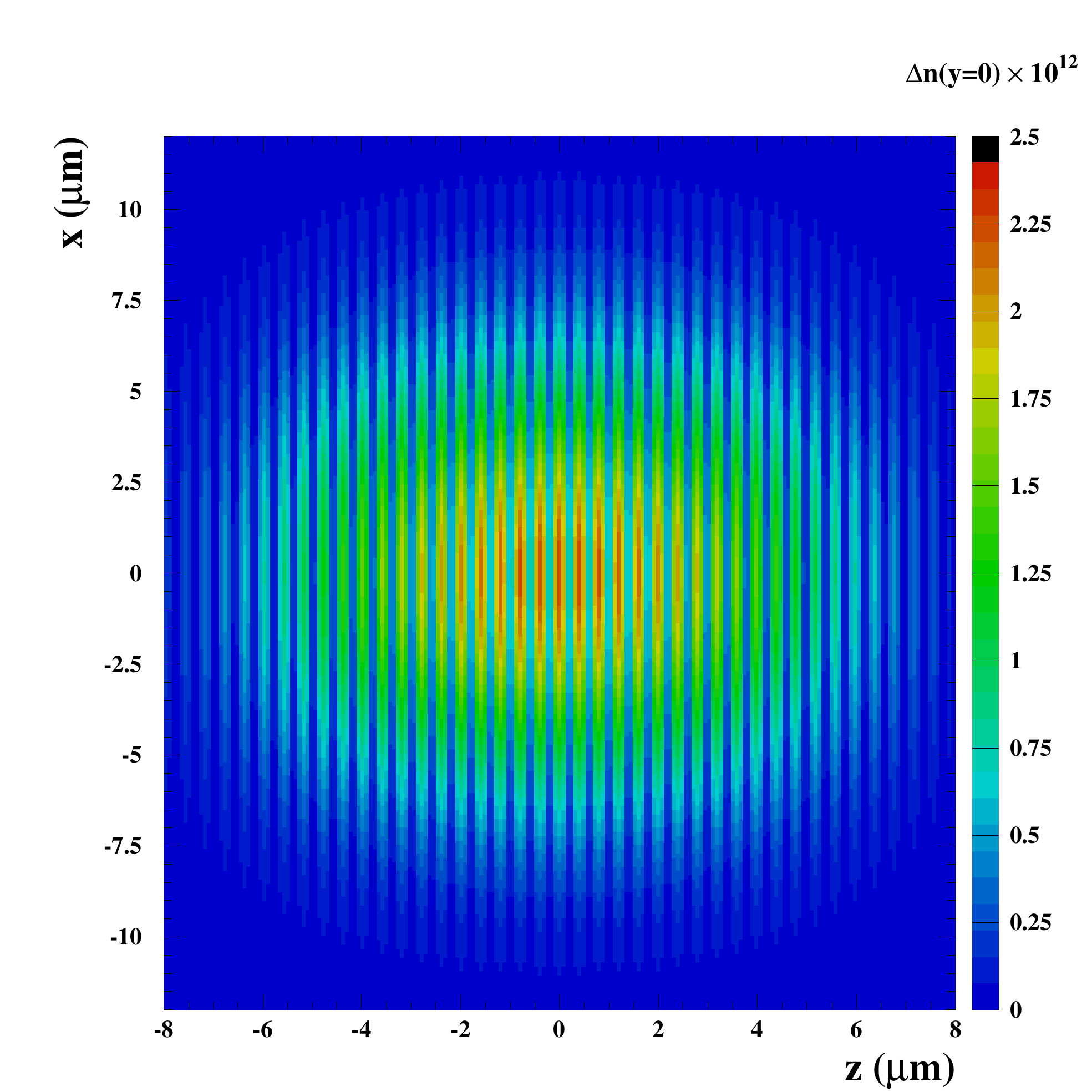}}
  \caption{The integrated index $\Delta n(x,y=0,z)$ ($\times 10^{12}$) calculated in the longitudinal plane $y=0$, as a function of the transverse coordinate $x$ and the longitudinal coordinate $z$. Computation is done with same conditions as for Fig.~\ref{fig:index-xy}.}
  \label{fig:index-xz}
\end{figure}

The integrated index $\Delta n(x,y,z=0)$ and refraction angle $\theta_r(x,y,z=0)$, both calculated in the transverse plane $z=0$, are presented as a function of $x$ and $y$ in Figures~\ref{fig:index-xy} and \ref{fig:angle}, respectively. 
At the maximum of intensity ($x=y=0$), the integrated index is about $2 \ 10^{-12}$ and at the maximum of the index gradient, the integrated refraction angle $\theta_{r,\mathrm{max}}$ is about 5~picorad. 
The integrated index $\Delta n(x,y,z=0)$ calculated in the longitudinal plane $y=0$  is also presented  as a function of the transverse coordinate $x$ and the longitudinal coordinate  $z$ in Figure~\ref{fig:index-xz}.

The refraction angle varies inversely as the cube of the minimum waist $w_0$ at the interaction area of the two pump pulses, and varies linearly with the total energy $I$ of the pump pulses, as
\begin{eqnarray}
\label{eq:angle-max}
\theta_{r,\mathrm{max}} = 2 \ 10^{-13} \times (\frac{w_0}{10 \mu\mathrm{m}})^{-3} \times \frac{I}{1 \mathrm{J}} \ \mathrm{ radians}   
\end{eqnarray}
In theory, the refraction angle is insensitive to the duration of the pump pulses since the refraction is integrated along all the longitudinal profile of the pulses. However, a waist of the order of 10~$\mu$m must be maintained in the interaction area. 
Therefore  the longitudinal size of the pulses must be smaller than few tens of micrometers in order to be unsensitive to the divergence of the beam along the interaction. This requires pulses shorter than few hundred femtoseconds.

\section{Measurement of the refraction with a Sagnac interferometer}

In our present proposal, the probe pulse is first focused in the interaction area of the two pump pulses and the refraction occurs at the focus. After refraction, the probe pulse is collimated again and analysed. Thus a refraction angle $\theta_r$ at the focus corresponds to a transverse shift $\delta r=f \times \theta_r$ after collimation, where $f$ is the focal length. 
The experimental challenge is therefore to detect an ultra small transverse displacement of the beam.

A transverse shift of a light beam is usually measured by using a quadrant photodetector and by recording the relative difference between the two measured intensities in each quadrant. In the case of a continuous intense laser beam, the high statistic of detected photons provides a very high sensitivity on the deviation angle of the light beam. For instance, a sensitivity of 0.1~picoradians at 1~kHz has been achieved by using an He-Ne continuous laser and a rotating mirror using an ultra low-noise locking amplifier synchronized with the rotation frequency of 1~kHz~\cite{lorrain}. 
However, with ultra short laser pulses, a photodetector would be highly saturated by the extremely high photon flux.

In order to overpass this limitation, we propose to measure the displacement of the probe beam by using a Sagnac interferometer.
Measurement of the deflection of a laser beam with a Sagnac interferometer has been already developped with continuous laser beams.
Dixon et al. have experimentally demonstrated that the use of a Sagnac interferometer allows for the ultra sensitive deflection measurement of a continuous laser beam~\cite{rochester}. An angular deflection of a mirror down to sub picoradians has been measured. A similar interferometer has also been developped more recently by Turner et al. who demonstrate a sensitivity of continuous laser beam deflection measurement better than 10~picoradians/$\sqrt{\mathrm{Hz}}$ between 10 and 200~Hz~\cite{seattle}.
But, to our knowledge, no deflection measurement has been performed in pulsed mode with a Sagnac interferometer. 

\begin{figure}\sidecaption
\resizebox{1.\hsize}{!}  {\includegraphics*{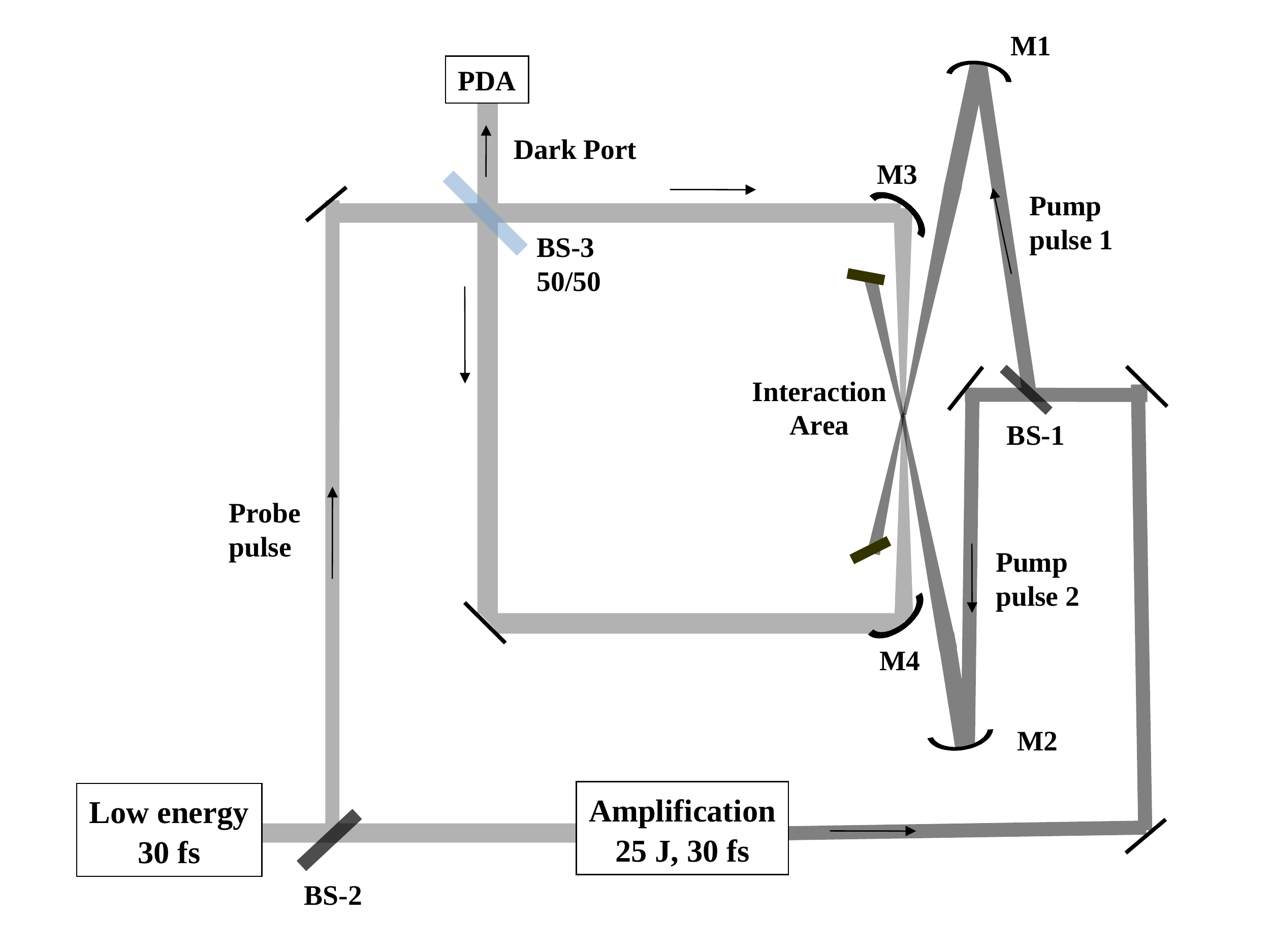}}
  \caption{Schematic view of a possible setup. Descriptions are given in the text.}
  \label{fig:setup}
\end{figure}

A possible setup is illustrated in Figure~\ref{fig:setup}.  
The intense laser pulse is divided by a beam splitter BS-1 in two counter-propagating pulses which interact at the focus of mirrors M1 and M2. A probe pulse, delivered by the beam splitter BS-2, is sent through a Sagnac interferometer via a 50/50 beam splitter BS-3. Two probe pulses, counter-rotating inside the Sagnac interferometer, are both focused by mirrors M3 and M4 into the interaction area of the two pump pulses. An array of photodetectors (PDA) is located in the dark output of the Sagnac interferometer. In the absence of interaction of the pump pulses, the two counter-rotating probe pulses are phase shifted by $\pi$ and no light is observed on the photodetector. The attenuation factor of the interferometer depends on the  symmetry of the beam-splitter BS-3 for the transmitted and reflected amplitudes.
When the two pump pulses interact, one of the probe pulses that crosses the interaction area at the same time, is refracted and is therefore transversely shifted at the dark output, while the second counter-rotating probe pulse crosses the interaction area much later and is therefore undisturbed. The small transverse displacement of one probe pulse relative to the second one generates an asymetry in amplitude at the dark output, and therefore a light signal appears.  
We also note that a vacuum of the order of $10^{-10}$~mbar is required in the interaction area of the two pump pulses, in order to be insensitive to nonlinear interactions in the residual gas.

There are specific advantages for the use of a Sagnac interferometer. 
First of all, the photo-detection is performed in the dark port at low intensity. 
Also, in the presence of beam pointing fluctuations, the interference pattern at the dark output is unmodified but only translated on the photo-detector plane. 
Finally a Sagnac interferometer is much less sensitive to mechanical fluctuations, and unsensitive to any linear group delay dispersion of the optical elements, since the two interfering beams travel along identical optical paths. 

We can estimate the sensitivity of the Sagnac interferometer by using a simplified model. We note $E(r)$ the transverse profile of the field amplitude for the unmodified probe pulse at the dark output of the interferometer, and $E(r+\delta)$ for the refracted pulse with the transverse displacement $\delta$. We note $\epsilon$ the asymetry between the transmitted and reflected intensity in the beam splitter. 
The transverse intensity profile $I(r)$ at the dark output is given by
\begin{eqnarray}
\nonumber I(r) & = & \left| \left(\frac{1+\epsilon}{\sqrt{2}}\right)^2 E(r+\delta) - \left(\frac{1-\epsilon}{\sqrt{2}}\right)^2  E(r) \right|^2 \\
\nonumber I(r) & \simeq & \frac{\delta^2}{4} \left(\frac{\partial E(r)}{\partial r}\right)^2  + 2 \epsilon \delta  E(r) \frac{\partial E(r)}{\partial r} + 4\epsilon^2 E^2(r)
\end{eqnarray}
Since $\delta^2 \ll \epsilon \delta$, the first term can be ignored and one gets
\begin{eqnarray}
\label{eq:intensity}
 I(r) = 2 \epsilon \delta \left|E(r)\right| \left|\frac{\partial E(r)}{\partial r}\right| + 4 \epsilon^2 E^2(r)
\end{eqnarray}
We now assume a simplified transverse beam profile, similar to a top-hat profile:
\begin{eqnarray}
\label{eq:2}
\nonumber E(r) = 
\left\{
\begin{array}{ll}
1 & \mbox{if $|r|\le \frac{\Delta r}{2}$} \\
\pm \frac{r}{a} + \frac{\Delta r}{2a} +1 \ & \mbox{if $\frac{\Delta r}{2} < |r| < \frac{\Delta r}{2} + a$} \\
0 & \mbox{if $|x| \ge \frac{\Delta r}{2} + a$}
\end{array}
\right.
\end{eqnarray}
In the region of interest, i.e. on the edge of the profile at $r_0 \simeq \Delta r/2 +a/2$, the intensity given in Eq.~\ref{eq:intensity} is
\begin{eqnarray}
\label{eq:4}
I(r_0) \simeq  \frac{\epsilon \delta}{a} + \epsilon^2
\end{eqnarray}
The first term $I_S=\epsilon \delta/a$ corresponds to the signal intensity  induced by the transverse displacement of the refracted pulse while the second term $I_B=\epsilon^2$ corresponds to the background intensity induced by the asymetry of the beam splitter. 
The signal over background ratio is 
\begin{eqnarray}
I_S/I_B \simeq \delta/(a \epsilon)
\end{eqnarray}
This ratio has to be compared to the relative uncertainty $\Delta I_B/I_B$ of the background intensity, which is given by the statistical error on the number of detected photons $N_B$ in the region of interest ($\Delta r/2 < |r| < \Delta r/2 + a$). Noting $N_0$ the number of photons of the initial probe pulse before entering the Sagnac interferometer, and $N_{shot}$ the number of repetitions of laser shots, then $N_B=a/\Delta r \times \epsilon^2  N_0 N_{shot} $, and the relative uncertainty of the background is
\begin{eqnarray}
\Delta I_B/I_B = 1/\sqrt{N_B} = (\sqrt{\Delta r/a})/(\epsilon \sqrt{N_0 N_{shot}})
\end{eqnarray}
The sensitivity of the measurement, given by the ratio of the signal over background fluctuation, is
\begin{eqnarray}
\label{eq:sensitivity}
\frac{I_S}{\Delta I_B} = \frac{\delta}{\sqrt{a \Delta r}} \sqrt{N_0 N_{shot}}
\end{eqnarray}
Assuming a beam diameter $\Delta r = 1$~mm, a width of the profile edge $a = \Delta r/10 = 100 \ \mathrm{\mu}$m, 
a repetition of laser shots $N_{shot}=10^4$ (corresponding to about one day of measurement with the LASERIX repetition rate of 0.1~Hz), 
and a transverse displacement $\delta = 10^{-11}$~m (corresponding to a refraction angle $\theta_r=10$~prad and a focal length $f=1$~m), 
the ratio is 
\begin{eqnarray}
\label{eq:result}
\frac{I_S}{\Delta I_B} = 3.10^{-6} \sqrt{N_0}
\end{eqnarray}
Therefore $N_0 = 10^{11}$ photons per pulse are required at the entrance of the interferometer in order to reach 1$\sigma$ sensitivity per day. Thus corresponds to $10^6$~W (25~nJ, 25~fs duration) per pulse. Highly Linear InGaAs photodiodes can handle a high power up to about 100~mW. By limiting the signal readout in the region of interest where the signal intensity is expected ($a/\Delta r \simeq 1/10$), an attenuation factor of the Sagnac interferometer of $\epsilon^2 \simeq 10^{-6}$ is required. 

To estimate the magnitude of the attenuation factor which can be achieved with femtosecond laser pulses, a Sagnac interferometer is being developped.

\section{Conclusion}

Several experimental tests using high intensity lasers have been proposed to observe electromagnetic nonlinearities in the vacuum, such as the diffraction or the reflection of intense laser pulses. 
In this paper we have shown the principles of a new and complementary method which consists in observing the refraction, i.e. the rotation of the waveplanes of a probe laser pulse crossing a transverse vacuum index gradient. The latter is produced by the interaction of two counter-propagating very intense and ultra short laser pulses, used as pump pulses. 
The transverse gradient of the vacuum refractive index is calculated and the refraction angle of the probe pulse crossing this index gradient is shown to be $0.2 \times (\frac{w_0}{10 \mu\mathrm{m}})^{-3} \times \frac{I}{1 \mathrm{J}}$~picoradians at the maximum. This corresponds, for the  most intense laser pulses attainable today by the LASERIX facility ($I = 25$~J) and with a minimum waist of $w_0 = 10$~$\mu$m at the focus, to  5~picoradians at the maximum. Although angular deviations as small as 0.1~picoradians have been detected with continuous laser beams with standard techniques using quadrant photodetectors, some limitations appear with ultra short and intense pulses. In order to overpass these limitations, we propose to use a Sagnac interferometer, for which the measurement of the refraction is performed in its dark port, thus allowing high intensity probe pulses. We estimate that the attenuation factor of the interferometer must be of the order of $10^{-6}$ in order to detect a signal larger than the background fluctuation after $10^4$ LASERIX shots (correponding to about 1~day of measurement).

\section*{Acknowledgments}

The authors thank J. Ha\"{\i}ssinski for fruitful discussions on this paper.
The authors also thank L-Acoustics Company for their financial support for building a Sagnac interferometer prototype.
We are also grateful to one of the referees who allowed us to clarify and complete our proposal.

\end{document}